\documentclass[twocolumn,prc,showpacs,floatfix]{revtex4}

\usepackage{amsmath,bm}
\usepackage{graphicx}

\begin{document}
\title{ Testing the theory of QGP-induced energy 
                                       loss at RHIC and the LHC }

\author{Ivan Vitev}

\affiliation{ Los Alamos National Laboratory, Theoretical Division and 
Physics Division, Mail Stop H846, Los Alamos, NM 87545, USA }

\date{\today}

\begin{abstract}

We compare an analytic model of jet quenching, based on the GLV 
non-Abelian energy loss formalism, to numerical results for the 
centrality dependent suppression of hadron cross sections 
in Au+Au and Cu+Cu collisions at RHIC. Simulations of neutral 
pion quenching versus the size of the colliding nuclear system 
are presented to high transverse  momentum $p_T$. At low and moderate 
$p_T$, we study the contribution of medium-induced gluon
bremsstrahlung to single inclusive hadron  production. In Pb+Pb 
collisions at the LHC, the redistribution of the lost energy is 
shown to play a critical role in yielding nuclear suppression 
that does not violate the participant scaling limit. 

\end{abstract}

\maketitle

\section{Introduction}

Strong suppression of single inclusive pions and charged 
hadrons  at large transverse momentum, as large as 
$p_T = 20$~GeV~\cite{Shimomura:2005en,Dunlop:2005xe}, 
is arguably one of the most fascinating phenomena 
from the heavy ion program at the Relativistic Heavy Ion 
Collider (RHIC). It signifies the transition from 
soft, collective and strongly 
model-dependent physics to the high $p_T$ or $E_T$ production 
of particles and jets that is well understood in terms of 
the perturbative Quantum Chromodynamics (pQCD) factorization
approach. Calibrated hard probes can thus be used to sample the 
properties of the  medium created in collisions of heavy 
nuclei~\cite{Gyulassy:2004vg}.  With capabilities to detect 
$p_T \sim 50$~GeV pions and $E_T \sim 500$~GeV jets with
good statistics, 
experiments at the Large Hadron Collider (LHC) will  
be able to critically test such perturbative calculations 
of hadron and  jet modification in the quark-gluon plasma 
(QGP) at a new energy frontier.

Following the discovery~\cite{Levai:2001dc} of jet quenching 
in Au+Au collisions and its verification through d+Au 
measurements~\cite{Adams:2003im} at RHIC, heavy ion theory has 
emphasized the need for a systematic study of the 
energy and system size dependence of leading particle 
attenuation~\cite{Gyulassy:2004vg,Vitev:2002pf}. Previous 
measurements have been limited to transverse momenta $p_T \leq 10$~GeV
and Au+Au collisions at RHIC~\cite{Adler:2003qi}.  More recent 
results up to $p_T \sim 20$~GeV~\cite{Shimomura:2005en,Dunlop:2005xe} 
in both Au+Au and Cu+Cu reactions were shown to be compatible with 
several theoretical estimates, emphasizing the final state QGP-induced 
suppression of jets~\cite{Vitev:2002pf,Adil:2004cn,Turbide:2005fk}. 
In this Letter, we compare for the first time 
analytic~\cite{Gyulassy:2004vg} and numerical~\cite{Vitev:2002pf} 
models of jet absorption in order to establish the role of 
sub-leading effects, such as the running of the strong coupling
constant $\alpha_s$ with the Debye screening scale 
$\mu \simeq g T$ in the plasma.  We present our predictions for 
$\pi^0$ quenching versus centrality, $p_T$ and $\sqrt{s}$, 
based on the Gyulassy-Levai-Vitev (GLV) approach to the 
medium-induced non-Abelian energy loss~\cite{Gyulassy:2000fs}.
We note that redistribution of the lost energy in 
small and moderate  $p_T$ hadrons plays a significant role 
in the transition from high $p_T$ suppression to low $p_T$ enhancement  
of the back-to back two particle correlations~\cite{Vitev:2005yg}. 
In this Letter we give results for single inclusive pions 
and identify the range of transverse momenta at RHIC and the LHC 
where the gluon feedback is  important.

In Section II we present a simple analytic model for the system
size, $N_{\rm part}$, dependence of jet quenching in nucleus-nucleus 
collisions. Section III contains select intermediate results from the 
GLV theory, including the evaluation of the mean gluon number 
$\langle N_g \rangle$ and fractional energy loss  
$\langle \Delta E / E \rangle$.  The derivation
of the radiative gluon contribution to small and moderate
$p_T$ hadron production is given in Section IV. Section V 
presents the calculated suppression of $\pi^0$ production 
in Cu+Cu and Au+Au collisions at RHIC and Pb+Pb collisions at the 
LHC. Conclusions are given in Section VI.

\section{ Analytic model of jet quenching }

While tomographic determination of the properties of the medium
created in nucleus-nucleus collisions can only be achieved through
detailed numerical simulations, it is useful to define a simplified 
analytic model which incorporates the essential features of jet 
quenching calculations. We first review the approach formulated
in \cite{Gyulassy:2004vg} and discuss its advantages and limitations. 

The final state medium-induced energy loss occurs after the 
hard partonic scattering $ab \rightarrow cd$ and before the 
fragmentation of the parent parton into a jet of 
colorless hadrons. We  consider for simplicity high  
$p_T > 5$~GeV hadron production at RHIC and will show later that 
the contribution of the induced bremsstrahlung in this region 
is small. There are two equivalent ways of implementing energy 
loss in the perturbative QCD hadron production formalism. 
The first one  associates $\Delta E$ with the kinematic 
modification 
of the momentum fraction $z = p_h^+ / p_c^+ $ in the fragmentation
function $D_{h/c}(z)$, leaving the hard parton production cross
section $d \sigma^{c} / dy d^2 p_{T_c}$ unmodified. The second 
approach reduces the jet cross section in the presence of the 
medium but leaves $D_{h/c}(z)$ unaltered. It can conveniently 
be implemented in an analytic model of QGP-induced leading hadron
suppression.
We take the underlying parton production cross section to 
be of power law type
\begin{equation}
\frac{d \sigma^{c}}{dyd^2p_{T_c}} = \frac{A}{(p_0 + p_{T_c})^{n}} 
\approx  \frac{A}{ p_{T_c}^n } \; , 
\;\;\;  {\rm if}\;\; p_{T_c} \gg p_0 \; ,
\label{pspectrum}
\end{equation}
where $n = n \, (y,p_{T_c},\sqrt{s})$ and 
$p_0 \sim 2.5$~GeV at RHIC. In a finite 
$p_T$ range, fixed rapidity $y$ and center of mass 
energy $\sqrt{s}$,  
a constant $n = \langle \, n \, (y,p_{T_c},\sqrt{s}) \, \rangle$ 
is a good approximation. At RHIC, in the region of 
$5 \; {\rm GeV} < p_{T_c} < 10 \; {\rm GeV}$,  the  
spectra scale roughly as $n_q = 7$,  $n_g = 8.4$. 
Fragmentation functions are convoluted with the partonic cross 
section as follows
\begin{eqnarray}
\frac{d \sigma^{h}}{dyd^2p_T} & = & 
\sum_c \int_{z_{\min} }^1 dz \; 
\frac{d \sigma^{c} (p_c = p_T/z) }{dyd^2p_{T_c}} \frac{1}{z^2} D_{h/_c}(z)
\nonumber \\
& \approx &   \sum_c
\frac{d \sigma^{c} (p_T/ \langle z \rangle) }{dyd^2p_{T_c}} 
 \frac{1}{ \langle z \rangle^2} D_{h/_c}(\langle z \rangle) \nonumber \\
& \approx &  \sum_c  \frac{A}{ p_{T_c}^n } \langle z \rangle^{(n-2)} 
D_{h/_c}(\langle z \rangle)
\label{hspectrum}
\end{eqnarray}
In Eq.~(\ref{hspectrum})  $z_{\min} = p_T / p_{T_{c}\, \max }$  
and the subsequent approximation is most reliable when
$z_{\min} \ll 1$. It should be noted that $ \langle z \rangle $ will
also depend on the partonic and hadronic species.

The second input to the analytic model comes from the radiative 
energy loss formalism~\cite{Gyulassy:2000fs}. To understand 
the system size dependence of jet quenching, we use the approximate 
GLV formula which relates $ \Delta E $  to the  size and the soft 
parton rapidity  density of the medium. For (1+1)D  Bjorken expansion, 
in the limit of large parton energy $2E/ \mu^2 L \gg 1$, we 
find~\cite{Gyulassy:2000gk}
\begin{equation}
 \frac{ \Delta E }{E}
\approx \frac{ 9 C_R \pi \alpha_s^3  }{4}
 \frac{1}{A_\perp} \frac{dN^{g}}{dy} \, L 
 \; \frac{1}{E} \, \ln   \frac{2 E}{\mu^2 L}  +  \cdots  \; .
\label{analyt-de}
\end{equation}
In Eq.~(\ref{analyt-de}) $L$ is the jet path length in the medium and 
$A_\perp$ is the transverse area.  $C_R = 4/3\; (3)$ for 
quarks (gluons), respectively, is the quadratic Casimir in the
fundamental (adjoint) representation of SU(3). Numerical simulations 
of  ${\Delta E }/{E}$  clearly indicate a weaker dependence  of the 
fractional energy loss on the jet energy than given in 
Eq.~(\ref{analyt-de}).

The key to  understanding the dependence of jet quenching on the 
heavy ion species and centrality is the effective atomic mass 
number, $A_{\rm eff}$, or 
the number of participants, $N_{\rm part}$, dependence of 
the characteristic plasma parameters in 
Eq.~(\ref{analyt-de})~\cite{Gyulassy:2004vg},  
\begin{eqnarray}
\label{A-dep-1}
&& {dN^g}/{dy} \; \propto \;  {dN^h}/{dy} \propto  A_{\rm eff} \;  
\propto \;  N_{\rm part} \; ,  \\
&& L \propto  A_{\rm eff}^{1/3} \; \propto \;  N_{\rm part}^{1/3} \;,  
\;\; A_{\perp} \propto A_{\rm eff}^{2/3} \; \propto \; N_{\rm part}^{2/3} \;. \quad
\label{A-dep}
\end{eqnarray}
Therefore, the {\em fractional} energy loss  scales approximately as   
\begin{equation}
\epsilon = { \Delta E }/{E} \; \propto \; A_{\rm eff}^{2/3} \; \propto \;   
N_{\rm part}^{2/3} \;,
\label{analyt-de-scal}
\end{equation}
up to logarithmic corrections from
Eq.~(\ref{analyt-de}).
If a parton loses this momentum fraction $\epsilon$ during its 
propagation 
in the medium to escape with momentum $p_{T_c}^{quench}$, immediately 
after the hard collision  $p_{T_c} = p_{T_c}^{quench} / 
( 1 - \epsilon )$. 
Noting the additional Jacobian $|d^2 p_{T_c}^{quench} 
/ d^2 p_{T_c}| = ( 1- \epsilon)^2 $, we find for the quenched 
hadronic spectrum per elementary $NN$ collision
\begin{eqnarray}
\frac{d \sigma^{h}_{quench}}{dyd^2p_T} & = &  \sum_c
\frac{d \sigma^{c} (p_T/(1-\epsilon)\langle z \rangle ) }{dyd^2p_{T_c}} 
\frac{1}{(1-\epsilon)^2\langle z \rangle^2} 
D_{h/_c}(\langle z \rangle)
\nonumber \\
& \approx & (1 - \epsilon_{\rm eff})^{n-2} 
\sum_c  \frac{A}{ p_{T_c}^n } \langle z \rangle^{(n-2)} 
D_{h/_c}(\langle z \rangle)  \; .
\label{hspectrum-quench} 
\end{eqnarray}
In Eq.~(\ref{hspectrum-quench}) $\epsilon_{\rm eff}$ is the 
average over all parton species and accounts for the color charge,
geometry and multi-gluon fluctuations.
From this result we can easily derive the system size 
dependence of the nuclear modification  factor
\begin{eqnarray} 
R_{AA} & = &  \frac{\sigma_{pp}^{in}}{N_{AA \; \rm col}} 
\frac{ d N^h_{AA}/dyd^2 p_T }
{ d \sigma^h  / dyd^2 p_T } \;\; (\rm exp.) \nonumber  \\[1ex]
 & \approx & 
\frac{ d \sigma^h_{quench}/dyd^2 p_T }
{ d \sigma^h  / dyd^2 p_T } \;\;  (\rm th.) \nonumber  \\
& = & (1-\epsilon_{\rm eff})^{n-2} = \left( 1 - \frac{k}{n-2} 
N_{\rm part}^{2/3}  \right)^{n-2} \;.
\label{raa-analyt} 
\end{eqnarray} 
In Eq.~(\ref{raa-analyt}) $k/(n-2)$ is the proportionality 
coefficient in Eq.~(\ref{analyt-de-scal}) which depends 
on the microscopic  properties of the medium but 
not on its size.  With $ n \gg 1 $ and experimentally 
measured and theoretically calculated suppression,
 $\sim 5$ fold in central Au+Au collisions at RHIC, 
$ \epsilon_{\rm eff} = N_{\rm part}^{2/3}\, k/(n-2) $ is small. 
We thus predict that the {\em logarithm} of nuclear 
suppression, Eq.~(\ref{raa-analyt}), has 
simple power law dependence on the system size
\begin{equation}
\ln R_{AA} = - k  N_{\rm part}^{2/3}  \; ,
\label{RAA-log}
\end{equation}
where the leading correction goes as 
$k^{2} N_{\rm part}^{4/3} / 2(n-2) $.

\begin{table}[!b]
\begin{center}
\begin{tabular}{||c||c|c|c|c|c|c|c|c|c||}
\hline 
  $Species$    &  $^9Be$
&     $^{16}O$  &  $^{28}Si$  &  
$^{32}S$  &  $^{56}Fe$  &  $^{64}Cu$  &  
 $^{197}Au$  &  $^{208}Pb$   &  $^{238}U$ \\
\hline \hline
  $b\; [fm]$  &  1 & 1  &  1.5 & 1.5  &  2  &  2 & 3  &  3 &  3 \\
\hline  
 $N_{part}^{2/3}$  &  5  &   8 &  12   &  14  & 20 & 22 
  &  48  & 50 & 55   \\ 
\hline
\end{tabular}
\end{center}
\caption{ \label{table-N23} Summary of the relevant quenching parameter 
$N_{\rm part}^{2/3}$ (rounded) at a fixed impact parameter $b$ for 
select heavy ion species. }
\end{table}

\begin{figure}[t!]
{\includegraphics[width=3.in,height=2.4in,angle=0]{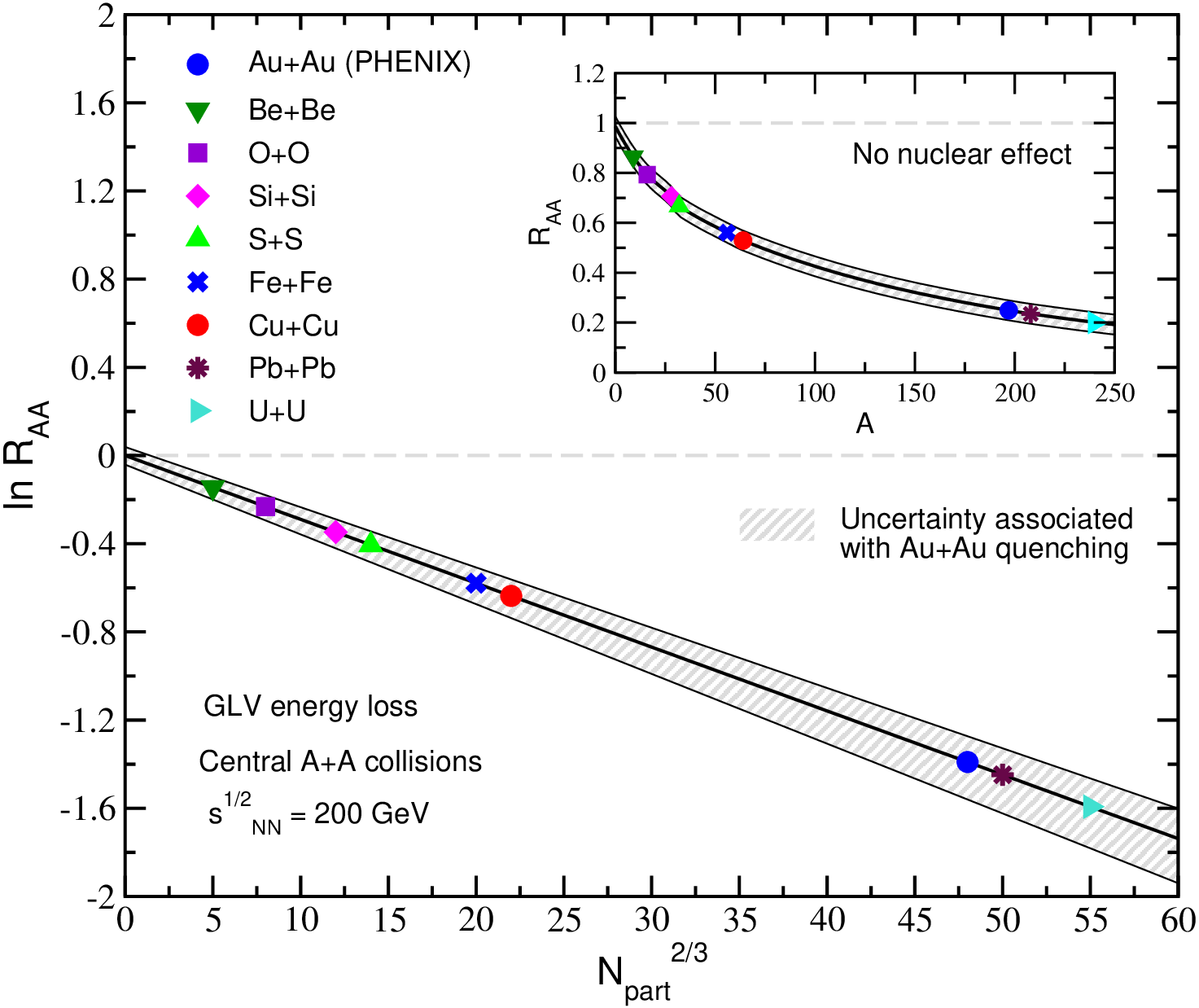}
\vspace*{1.cm} }

\vspace*{0cm}
\includegraphics[width=3.in,height=2.2in,angle=0]{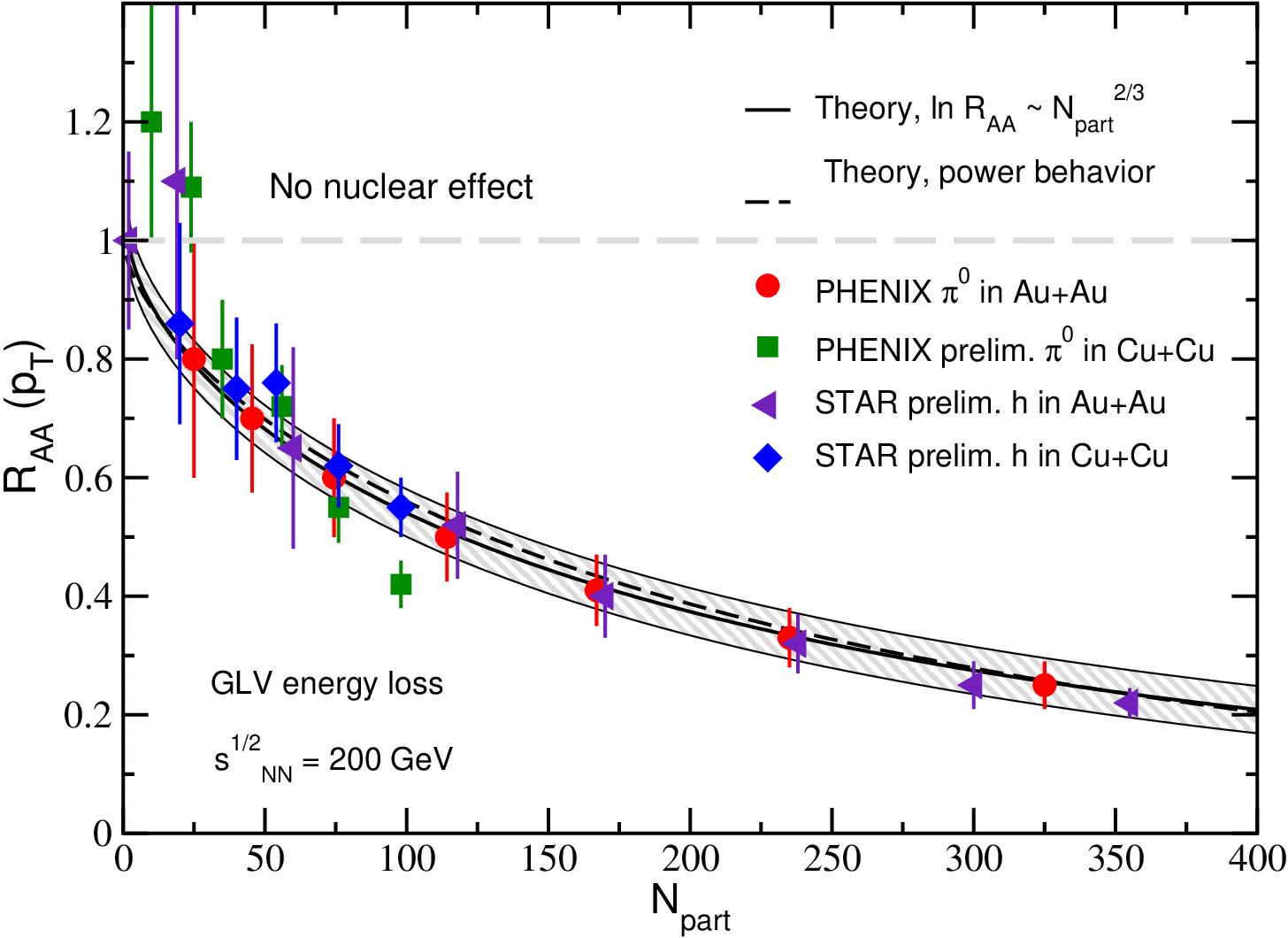}
\caption{ Top panel: the predicted linear dependence of jet 
quenching in natural variables, $\ln R_{AA}$ versus $N_{\rm part}^{2/3}\,$,
for central collisions of $^9$Be, $^{16}$O, $^{28}$Si,
$^{32}$S, $^{56}$Fe, $^{64}$Cu, $^{208}$Pb and  $^{238}$U.
Bottom panel: comparison of the analytic $R_{AA}$ to PHENIX and 
STAR experimental data in Au+Au and Cu+Cu collisions. }
\label{analytic}
\end{figure}

We focus on the most central nuclear collisions and use
impact parameters $b=1 - 3$~fm depending on the atomic mass $A$. 
An optical Glauber model calculation is used to evaluate 
$N_{\rm part}^{2/3}$  with $\sigma^{in}_{pp} = 42$~mb and 
a Wood-Saxon nuclear density with results given in 
Table~\ref{table-N23}. The analytic prediction for the 
system size dependence of jet quenching
is shown in Fig.~\ref{analytic}. It is fixed by the magnitude 
of the suppression established in central Au+Au
collisions~\cite{Adler:2003qi} and consistent with existing 
simulations~\cite{Gyulassy:2000gk}. From this analysis
we expect a factor $\sim 2$ suppression in central Cu+Cu 
collisions.  Comparison to the preliminary PHENIX
$p_T > 7$~GeV data~\cite{Shimomura:2005en}
and STAR  $p_T > 6$~GeV  data~\cite{Dunlop:2005xe} is also 
shown in the bottom 
panel of Fig.~\ref{analytic} with good agreement within 
the experimental uncertainties. Dashed and solid lines 
illustrate the difference between Eqs.~(\ref{raa-analyt}) 
and  (\ref{RAA-log}) when normalized to 
the same suppression in central Au+Au collisions.

The main advantage of the GLV analytic model is the ability to 
provide guidance on the magnitude of the QGP-induced jet quenching
versus centrality and address a large body of experimental 
data. Its limitations include a fixed coupling constant 
$\alpha_s$, implementation of only the mean, 
though suitably reduced, energy  loss $\Delta E$
and the inability to incorporate additional nuclear effects, 
such as the Cronin multiple scattering~\cite{Adams:2003im} 
and nuclear shadowing~\cite{Eskola:1998df}. It also relies 
on a reference numerical calculation in central A+A 
collisions in the same $p_T$ and $y$ range as well as
$\sqrt{s}$~\cite{Gyulassy:2000gk}. The deviation of $dN^g/dy$ 
from the exact participant scaling in Eq.~(\ref{A-dep-1}) may 
lead to less quenching and improved agreement with the data 
in peripheral reactions, but is here neglected.

\section{Numerical evaluation of the QGP-induced energy loss}

The solution for the differential in energy ($\omega$) and transverse 
momentum (${\bf k}$) spectrum of medium induced gluon 
radiation has been obtained order by order in the correlations 
between the multiple scattering centers in nuclear matter using 
the reaction operator approach~\cite{Gyulassy:2000fs} 
\begin{eqnarray}
 \omega \frac{dN_g}{d\omega \, d^2 {\bf k}}   & \approx &
  \sum\limits_{n=1}^{\infty}  \frac{C_R \alpha_s}{\pi^2} 
 \; \prod_{i=1}^n \;\int_0^{L-\sum_{a=1}^{i-1} \Delta z_a } 
 \frac{d \Delta z_i }{\lambda_g(i)} \nonumber \\[1.ex] 
&&  \times  \int   d^2{\bf q}_{i} \, 
\left[  \sigma_{el}^{-1}(i)\frac{d \sigma_{el}(i)}{d^2 {\bf q}_i}
  - \delta^2({\bf q}_{i}) \right]  \,  \nonumber \\[1.ex] 
&& \times 
\left(  -2\,{\bf C}_{(1, \cdots ,n)} \cdot 
\sum_{m=1}^n {\bf B}_{(m+1, \cdots ,n)(m, \cdots, n)}  
\right.  \nonumber \\[1.ex] 
&&   \times \left[ \; \cos \left (
\, \sum_{k=2}^m \omega_{(k,\cdots,n)} \Delta z_k \right) 
\right.  \nonumber \\[1.ex] 
&&  \left. \left.  \hspace*{.2cm} 
-   \cos \left (\, \sum_{k=1}^m \omega_{(k,\cdots,n)} \Delta z_k \right)
\right]\; \right) \;. \quad \qquad  
\label{difdistro} 
\end{eqnarray}
Here  ${\bf q}_i$ are  
the momentum transfers from the medium, 
distributed according to a normalized elastic differential cross section  
$\sigma_{el}(i)^{-1} {d \sigma_{el}(i)}/ {d^2 {\bf q}_i}$,  and 
$ \Delta z_k = z_k - z_{k-1} $ are
the separations of the subsequent scattering centers.  
In Eq.~(\ref{difdistro})  
the color current propagators and inverse formation times are denoted by
\begin{eqnarray}
{\bf C}_{(m, \cdots ,n)} &=&  \frac{1}{2} \nabla_{{\bf k}} 
\ln \, ({\bf k} - {\bf q}_m - \cdots  - {\bf q}_n )^2 \;\;, \nonumber \\  
{\bf B}_{(m+1, \cdots ,n)(m, \cdots, n)} &=&  
{\bf C}_{(m+1, \cdots ,n)} - {\bf C}_{(m, \cdots ,n)} \;\;, \nonumber \\  
\omega_{(m,\cdots,n)}  & = & 
\frac{({\bf k} - {\bf q}_m - \cdots  - {\bf q}_n )^2}{2 \omega}  \;. 
\label{props}
\end{eqnarray}

Numerical results have been obtained to third order in the opacity, 
$ \langle L / \lambda_g \rangle $, with all correlations 
of up to four scattering centers, including the initial hard 
interaction. To speed up the evaluation of the squared amplitudes,
the oscillating quantum coherence phases between the points of
interaction were converted to Lorentzians using an exponentially 
falling geometry with no sharp edges. The requirement that
the mean locations $z_k$, $k= 1 \cdots n$ of $n$ scatterers 
are the same as the one for a uniform soft parton distribution 
fixes the parameter $L_{e}$ of such geometry~\cite{Gyulassy:2000fs}:
\begin{equation}    
\langle z_0 - z_k  \rangle = k \frac{L}{n+1} \quad \rightarrow
\quad L_{e} = \frac{L}{n+1} \;.
\label{location}
\end{equation}       
Bjorken (1+1)D expansion of the plasma is accounted for as follows: 
\begin{eqnarray}    
&&  \!\!\!  \rho(z_k) = \rho (z_0) \frac{z_0}{z_k} \;,  \nonumber \\
&& \!\!\!   \mu(z_k) = \mu (z_0) \left( \frac{z_0}{z_k} \right)^{1/3}, 
\;\;  
\lambda_g(z_k)  =  \lambda_g (z_0) \left( \frac{z_k}{z_0} \right)^{1/3} 
, \qquad \label{bjorken}
\end{eqnarray}  
and the kinematic constraints, 
\begin{eqnarray}
&& \mu / Q \leq x = k^+/E^+ \approx \omega /E  \leq 1 \; , \nonumber \\  
&& \mu \leq  |{\bf k}| \leq \sqrt{Q^2} \min (x, 1-x) \;, \qquad \qquad
\label{bounds}
\end{eqnarray}
have been incorporated for consistency with our previous 
work~\cite{Gyulassy:2000gk}. It was recently shown that for physical 
on-shell final state gluons the medium induced radiation is infrared 
and collinear safe~\cite{Vitev:2005yg}. This allows 
relaxation of the  $\mu/Q \leq  x $, $\mu \leq  |{\bf k}| $ 
constraints in the future, though it 
should be noted that the Debye screening scale still controls the 
small ${\bf k}$ and $\omega$ cancellation between the single
and double Born diagrams in the opacity 
expansion~\cite{Gyulassy:2000fs}.

To evaluate the effect of multiple gluon emission and arrive at a 
probabilistic distribution $P(\epsilon)$ for the fractional energy 
loss $\epsilon = \Delta E / E = \sum_{i = 1}^n \epsilon_i$, 
$\epsilon_i = \omega_i /E$, we are
motivated by an independent Poisson gluon emission ansatz, 
but incorporate kinematic constraints~\cite{Arleo:2002kh}:
\begin{eqnarray} 
&& P(\epsilon ) = \sum_{n = 0}^\infty  P_n(\epsilon )  \;, 
\qquad P_0(\epsilon) = e^{-\langle N_g \rangle} \delta(\epsilon) \; ,
\nonumber \\
&& P_n (\epsilon ) = \frac{1}{n} \int_0^\epsilon d \epsilon^\prime \; 
P_{n-1}(\epsilon - \epsilon^\prime) 
\frac{dN_g}{d\epsilon^\prime} (\epsilon^\prime = \omega/E) \;.  \qquad  
\label{poiss} 
\end{eqnarray}
We normalize this probability density to unity and Eq.~(\ref{poiss})
ensures that at every step energy is conserved. As a consequence,
for small jet energies and large $\Delta E / E$ the gluon 
distribution is distinctly non-Poisson. We evaluate the mean
energy loss as follows:
\begin{eqnarray} 
&& \int_0^1 d \epsilon \; P( \epsilon ) = 1\;, \quad 
 \int_0^1 d \epsilon \; \epsilon \, P( \epsilon ) = 
\left\langle  \frac{ \Delta E }{ E } \right\rangle \;. \quad 
\label{meane} 
\end{eqnarray}

\begin{figure}[t!]
\includegraphics[width=3.3in,angle=0]{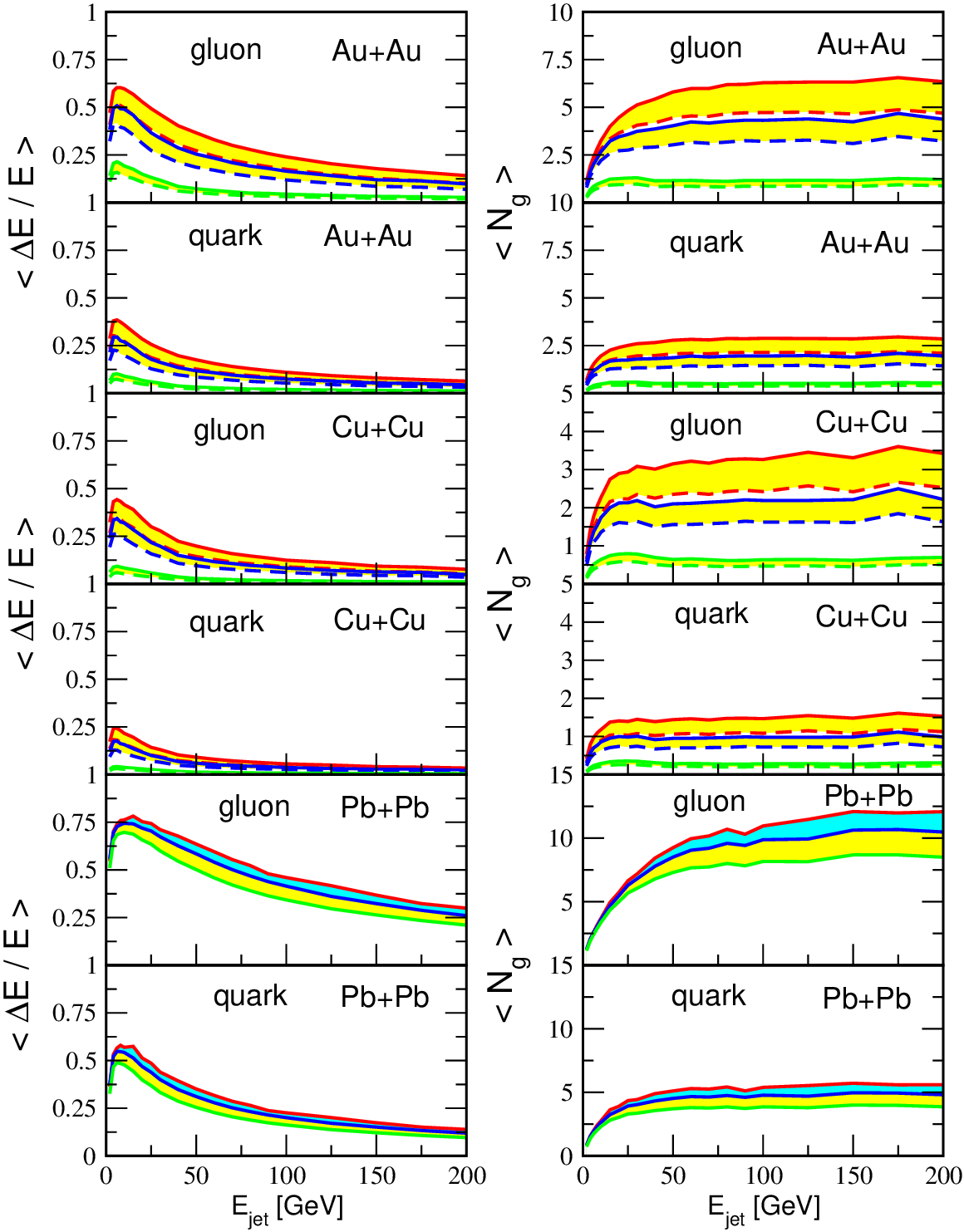}
\caption{ Left panel: mean energy loss  
for Au+Au and Cu+Cu soft parton densities at RHIC corresponding
to centralities given in Table~\ref{table-AuCu} and 
Pb+Pb soft parton densities 
at the LHC corresponding to central collisions 
versus the jet energy. Right panel: mean 
gluon number versus $E_{jet}$ for the same heavy ion 
systems.}  
\label{ElossNg}
\end{figure}

\begin{table}[!b]
\begin{center}
\begin{tabular}{||c||c|c|c||}
\hline 
 \;  $Centrality$   \;   &  $0 - 10\% $
&   $20-30 \% $  & \;  $60-80\%$  \;  \\
\hline \hline
 $N_{\rm part} $  &  328  &  167  &  21 \\
\hline
$dN^g/dy$ & \;  800 - 1175 \;  & \; 410 - 600 \; & \; 50 - 75  \;  \\
\hline
  $L$ [fm]  &  6 & 4.8 & 2.4  \\
\hline  
 $A_{\rm eff}$  &  197  &  99  &  12   \\ 
\hline
\end{tabular}

\vspace*{.4cm}

\begin{tabular}{||c||c|c|c||}
\hline 
 \;  $Centrality$   \;   &  $0 - 10\% $
&   $20-30 \% $  & \;  $60-80\%$  \;  \\
\hline \hline
 $N_{\rm part} $  &  103  &  55  &  9 \\
\hline
$dN^g/dy$ & \;  255 - 370 \;  & \; 135 - 195 \; & \; 20 - 30  \;  \\
\hline
  $L$ [fm]  &  4.1 & 3.3 & 1.8  \\
\hline  
 $A_{\rm eff}$  &  64  &  34  & 6   \\ 
\hline
\end{tabular}
\end{center}
\caption{\label{table-AuCu} Estimated $dN^g/dy$, $L$ and 
$A_{\rm eff}$ versus $N_{\rm part}$ for central, semi-central
and peripheral Au+Au (top table) and Cu+Cu (bottom table) 
at RHIC. } 
\end{table}

In Ref.~\cite{Gyulassy:2000gk} we considered jet production
following the binary collision density $T_{AA}(b)$ in central 
Au+Au reactions  at $\sqrt{s}=200$~GeV. In an elementary 
hard interaction inclusive jets are distributed uniformly
in azimuth relative to the reaction plane. We calculated 
 $\langle \langle \, \Delta E \,  
\rangle \rangle_{\rm geom.} $ using the line integral, 
Eq. (\ref{difdistro}),  
through the (1+1)D Bjorken expanding medium 
density by correctly weighing the
amount of lost energy with the jet production rate.  
Cylindrical geometry with radius $L=6$~fm that gives 
the same mean energy loss $\langle \,\Delta E \, \rangle $  
for uniform initial soft parton rapidity density was 
then  constrained. Using Eqs.~(\ref{A-dep-1}) and
(\ref{A-dep}) we can determine the effective length $L$, 
transverse area $A_\perp = \pi L^2$, gluon rapidity density 
$dN^g/dy$ and effective atomic mass $A_{\rm eff}$ 
for shadowing applications~\cite{Eskola:1998df} in 
interacting heavy ion systems of different size.
Table~\ref{table-AuCu} summarizes the parameters used in 
our calculation of central, mid-central and peripheral Au+Au and 
Cu+Cu collisions at $\sqrt{s}=200$~GeV at RHIC. For central
collisions with $\sqrt{s}=5.5$~TeV at the LHC, we use 
$dN^g/dy = 2000, \, 3000$ and $4000$ to test the sensitivity 
of jet quenching to the QGP properties.

The calculated fractional energy loss and mean gluon number
for quark and gluon jets versus their energy for the centralities
and densities discussed above are shown in 
the left and right panels of Fig.~\ref{ElossNg}. Only in the 
limit $E_{jet} \rightarrow \infty$, $ \Delta E / E 
\rightarrow 0$ does the energy loss for quarks and gluons
approach the naive ratio  $\Delta E^g / \Delta E^q = C_A/C_F = 9/4$.
For large fractional energy losses this ratio is determined by 
the $\Delta E \leq E $ constraint. In this regime no simple 
scaling arguments related to the properties of the dense nuclear 
matter and the color charge are applicable.
It should be noted that at high $E_{jet}$ the fractional energy 
loss is not large even at the LHC.

In our calculation the  strong coupling constant 
$\alpha_s$ is not used as a free parameter but evaluated 
at the typical scale $\mu$ in the elastic scattering cross
section  $\sigma_{el}(i)$ and the gluon mean free path
$\lambda_g(i)=1/\sigma_{el}(i) \rho(i)$. At the radiation 
vertex, $\alpha_s({\bf k}^2)$ is also sensitive to the increase 
of the temperature or density of the medium, 
$\rho \propto T^3$, with the increase of $dN^g/dy$ at 
a fixed transverse area $A_\perp$. The effects
described here lead to sub-linear dependence of the energy
loss on $dN^g/dy$. For the LHC example, given in 
Fig.~\ref{ElossNg}, this can be a 50\% correction for
large $E_{jet}$ and even more significant at low $E_{jet}$
when compared to Eq.~(\ref{analyt-de}).   
Conversely, at low parton densities in peripheral collisions
the energy loss will be larger than naively expected.

Another important point, seen in the right panel of 
Fig.~\ref{ElossNg}, is that except for very low jet energies
the mean gluon number $ \langle N_g \rangle $ is not small
($\ll 1$) and the probability of not radiating gluons, 
$P_0 = \exp(- \langle N_g \rangle )$,  is never large. 
Full results for $ P(\epsilon) $  were shown 
in~\cite{Adil:2004cn,Arleo:2002kh}. We find that the
probability density {\em does not} approximate 
$ a \delta(\epsilon) + (1-a) \delta(1-\epsilon) $, 
$ a < 1 $. The latter ansatz yields $ R_{AA}(p_T) = a $ 
independent on the collision energy or $p_T$.
Instead, $P(\epsilon)$ is much more uniformly distributed in
the interval $[0,1]$ and our calculations retain sensitivity
to the local slope and the parton species contribution 
to the differential inclusive hadron production cross section, 
see  Eq.~(\ref{raa-analyt}).

\begin{figure}[!t]
\includegraphics[width=3.2in,height=2.1in,angle=0]{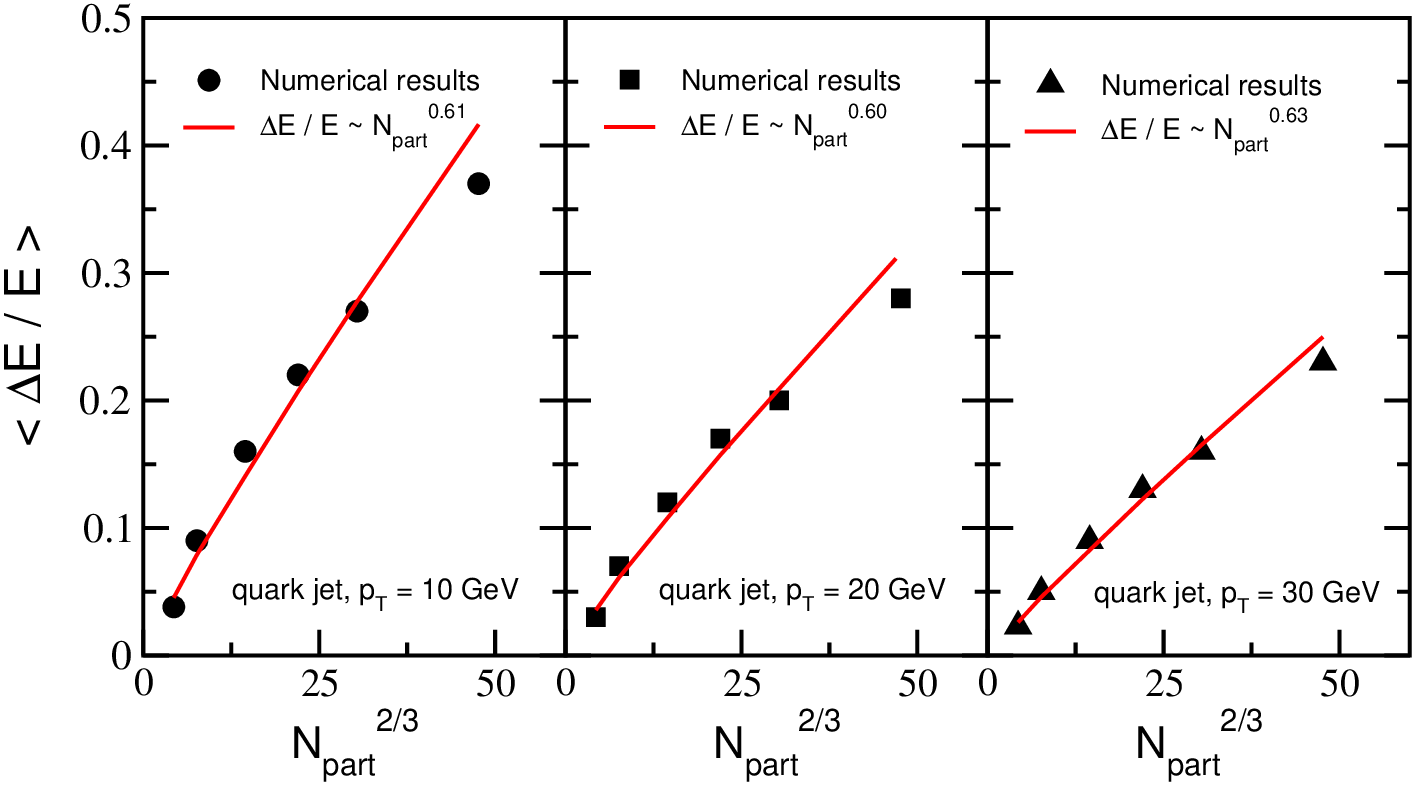}
\caption{ $N_{\rm part}$ dependence of the fractional energy 
loss $ \Delta E / E $ for $E_{jet} = 10, 20$ and $30$~GeV quark
jets in $\sqrt{s}  = 200$~GeV nucleus-nucleus collisions. 
Power law fits yield exponents $n = 0.61, 0.60$ and $0.63$,
respectively. } 
\label{cros-scal}
\end{figure}

We finally note that multi-gluon fluctuations, given by  
Eq.~(\ref{poiss}), reduce the jet quenching effect 
relative to the application of the mean 
$ \langle \; \Delta E / E \;  \rangle $ 
shown in Fig.~\ref{ElossNg}~\cite{Arleo:2002kh}. This can 
be seen by comparing the $\epsilon_{\rm eff} \approx 0.2$ 
in central Au+Au reactions, obtained from Eq.~(\ref{raa-analyt}),   
to the fractional energy loss for 10 GeV quark jets, 
$\epsilon_{\rm eff} < \langle \; \Delta E / E \;  \rangle $. 
To investigate the scaling of energy loss with $N_{\rm part}$
we select quark jets of $E_{jet} =10, 20$ and $30$~GeV at
midrapidity at RHIC and $N_{\rm part} = 9, 21, 55, 103, 167$ 
and  $328$. These cover fractional energy losses 
$ 0  < \langle \; \Delta E / E \;  \rangle < 0.4 $ with 
numerical results shown in Fig.~\ref{cros-scal}. Power law 
fits, also plotted, give  exponents $ n = 0.60 - 0.63 $  
that are not very different from the naive  $n = 2/3$ 
expectation from Eq.~(\ref{analyt-de-scal}) that was used in 
Fig.~\ref{analytic}. We conclude that the deviation 
between the calculated energy loss with kinematic constraints 
and running strong coupling constants and its asymptotic
fixed $\alpha_s$ analytic behavior is smaller 
when the variation of the energy loss 
is associated with a change in the system size rather than a  
large change in the density of the medium alone.

\section{ Nuclear effects on inclusive hadron production }

\begin{figure}
\includegraphics[width=3.2in,height=2.6in,angle=0]{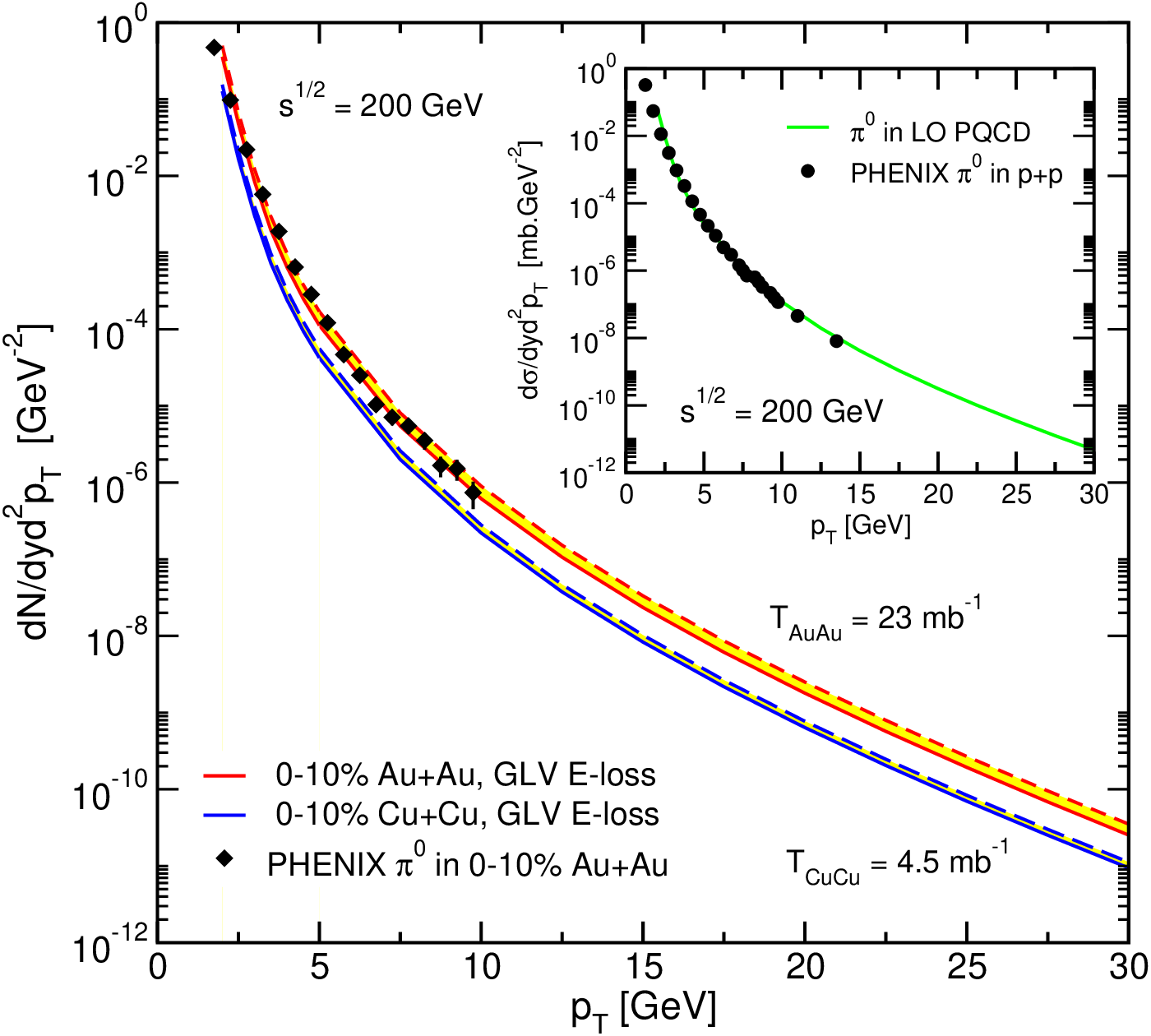}
\caption{ The predicted invariant multiplicity distribution
of neutral pions in central Au+Au collisions at 
$\sqrt{s} = 200\; GeV$ for 
medium density $dN^g/dy = 800 - 1175 $ and $T_{AuAu}=23\; mb^{-1}$.
The same calculation for Cu+Cu collisions for 
medium density $dN^g/dy = 255 - 370 $ and $T_{CuCu} = 4.5\; mb^{-1}$.
The insert shows the cross section for $\pi^0$ production 
in p+p collisions to LO pQCD.  Data is from PHENIX~\cite{Adler:2003qi}.}
\label{abs-cross}
\end{figure}

The lowest order perturbative QCD cross section for 
single inclusive hadrons in nucleon-nucleon (NN) 
reactions, including non-vanishing transverse momentum 
${\bf k}_{a,b}$ distributions of the incoming partons, 
is given by  
\begin{eqnarray}
\frac{ d\sigma^{h}_{NN} }{ dy  d^2p_{T}  }  & = & K
\sum_{abcd}   \int_{x_{a\, \min}}^1  \int_{x_{b\, \min}}^1 
d x_a d x_b 
\int d^2 {\bf k}_a  d^2 {\bf k}_b \;   \nonumber \\
& \times&   f({\bf k}_{a}) f({\bf k}_{b})
\, { \phi_{a/N}({x}_a, \mu_f) \phi_{b/N}({x}_b,\mu_f) }
\nonumber \\
& \times & \frac{1}{ \pi z_c } 
\frac{ d \sigma^{ ab \rightarrow cd } }{ d\hat{t} }  
D_{h/c}(z_c,\mu_f) \;.
\label{single}
\end{eqnarray}
Here $z_c = p_{T}/p_{T_c}$, $x_{a,b} = p_{a,b}^+/P_{a,b}^+$.
In our notation $\phi_{a,b/N}({x}_b,\mu_f)$ are the 
parton distribution functions (PDFs),  $D_{h/c}(z_c,\mu_f)$ 
are the fragmentation functions (FFs) and we have chosen 
the factorization/fragmentation, and renormalization 
scales $\mu_f =\mu_r = {p_{T_c}}$. The distribution of non-zero 
transverse momenta of the incoming partons is parametrized
as follows  
\begin{equation}
 f({\bf k}_{a,b}) = \frac{1}{ \pi \langle k_{a,b}^2 \rangle }
\exp (-{\bf k}_{a,b}^2/ \langle k_{a,b}^2 \rangle ) \; .  
\label{gauss}
\end{equation}
The physical requirement for hard partonic scattering is 
ensured by  $k_{a,b} < x_{a,b} \sqrt{s}$ and  $K = 1.5$ is 
a phenomenological K-factor at RHIC.  For further 
details see~\cite{Vitev:2002pf,Adil:2004cn}.  
Comparison to the PHENIX measurement of $\pi^0$ production in
$\sqrt{s}=200$~GeV p+p collisions at RHIC is shown in the 
insert of Fig.~\ref{abs-cross}.

Nuclear effects can be incorporated in the pQCD formalism,
Eq.~(\ref{single}), and fall in two categories: 
medium-induced kinematic modifications to the perturbative 
formulas and possibly universal modifications to the PDFs 
and FFs. An example of the latter are parameterizations of
nuclear shadowing $S_{q,g/A}(x,\mu_f)$~\cite{Eskola:1998df} 
included in this calculation via
\begin{eqnarray}
\frac{1}{A} \, \phi_{q,g/A}(x,\mu_f) & = &  
 \left( \frac{Z}{A} 
S_{q,g/A}(x,\mu_f)  \phi_{q,g/p}(x,\mu_f) \right. 
\nonumber \\
& + & \left. 
\frac{N}{A} S_{q,g/A}(x,\mu_f)  \phi_{q,g/n}(x,\mu_f) \right) \;.         
\qquad
\label{isoemc}
\end{eqnarray}
In future work we will combine dynamical calculations of
coherent nuclear enhanced power corrections with other 
elastic and inelastic effects in nuclear 
matter~\cite{Eskola:1998df}. 
Transverse momentum broadening of the incoming partons, 
leads to enhancement of the differential particle 
distributions at $p_T \sim$~few GeV~\cite{Adams:2003im} 
and can be accounted for in  Eqs.~(\ref{single}), 
(\ref{gauss}) as follows: 
\begin{eqnarray}
\langle k_{a,b}^2 \rangle = \langle k_{a,b}^2 \rangle_{vac} 
+  \langle k_{a,b}^2 \rangle_{med} \;.
\label{cronin}
\end{eqnarray}
Here $\langle k_{a,b}^2 \rangle_{med} =  
( 2  \mu^2 L_{a,b} / \lambda_{a,b} ) \xi $
and the typical momentum transfers squared $\mu^2$, mean free 
paths  $\lambda_{a,b}$ 
and parton propagation lengths $L_{a,b}$ refer to cold nuclear 
matter~\cite{Adams:2003im,Vitev:2002pf}.

In the QGP, final state energy loss is the dominant effect
that alters the single and double inclusive hadron production
cross sections~\cite{Adams:2003im,Vitev:2002pf}. 
Application to the attenuation of leading hadrons as a 
{\em kinematic } modification of the momentum fraction $z$
in the FFs $ D_{h/c}(z) $ is  considered 
standard~\cite{Vitev:2002pf,Adil:2004cn,Turbide:2005fk}.   
The redistribution of the lost energy 
in soft and moderate $p_T$ hadrons was only recently 
derived in the pQCD approach, first for back-to-back two 
particle correlations~\cite{Vitev:2005yg}. The established
dramatic transition from the high $p_T$ factor of four suppression
($R_{AA} \sim 0.25$) to the low $p_T$ factor of two enhancement
($R_{AA} \sim 2$) makes it imperative to study this effect 
for single inclusive particle production. One possibility 
is that the fraction of the hadrons from the bremsstrahlung 
gluons is negligible or small over the full 
accessible $p_T$ range. At the other extreme, a very large 
fraction may compromise the current jet quenching 
phenomenology, leading to $R_{AA} \sim 1$ at moderate 
transverse momenta even in dense matter.

With this motivation, we first give results for the 
modification of inclusive hadron production  from final state
radiative energy loss. It can be represented as:
\begin{eqnarray}
D_{h/c} (z) & \Rightarrow & \int_0^{1-z} d\epsilon \; P(\epsilon)  \; 
\frac{1}{1-\epsilon} D_{h/c} \left( \frac{z}{1-\epsilon} \right) 
\nonumber \\  && 
+  \,  \int_z^{1} d\epsilon \; \frac{dN^g}{d \epsilon}(\epsilon) \;  
 \frac{1}{ \epsilon }  D_{h/g} \left( \frac{z}{\epsilon} \right) \;\; .
\label{nucmod-1}
\end{eqnarray}
Here $P(\epsilon)$ is calculated from Eqs.~(\ref{difdistro}), 
(\ref{poiss}) and $ d N_g / d \epsilon$ is the distribution 
of the average gluons versus $\epsilon = \omega/E$ so that
\begin{equation}
\! \!\! \int_0^1 d\epsilon \; \frac{dN^g}{d \epsilon}(\epsilon) 
= \langle N_g  \rangle  \;, \;  
\int_0^1 d\epsilon \; \epsilon \frac{dN^g}{d \epsilon}(\epsilon) 
= \left\langle \frac{\Delta E}{E} \right\rangle \;.  
\label{avgue}
\end{equation}
It is easy to verify the momentum sum rule for all hadronic 
fragments from the attenuated jet and the radiative gluons. With 
appropriate changes of variables 
\begin{eqnarray}
&& \!\!\!\!\! \!\!\! 
  \sum_h \int_0^1 dz \; z  D_{h/c} (z)  \;   \Rightarrow  \nonumber \\
&& \quad \int_0^1 d \epsilon \;  (1-\epsilon )  P(\epsilon)  
  \sum_h \int_0^{1-\epsilon} \frac{dz}{1-\epsilon} \; 
\frac{z}{1 - \epsilon} D_{h/c} \left( \frac{z}{1-\epsilon} \right) 
\nonumber \\
&&\quad  + \int_0^1 d\epsilon \;  \epsilon  \, 
\frac{dN^g}{d \epsilon}(\epsilon) \;  
 \sum_h \int_0^\epsilon \frac{dz}{\epsilon} \; 
\frac{z}{\epsilon} D_{h/g} \left( \frac{z}{\epsilon} \right) 
\nonumber \\[1ex]
&& \quad = 1 - \langle \epsilon \rangle  
+  \langle \epsilon \rangle  = 1
\;\; .
\end{eqnarray}

Figure~(\ref{abs-cross}) shows the invariant $\pi^0$ multiplicity
in central Au+Au and Cu+Cu reactions. Data is from 
PHENIX~\cite{Adler:2003qi}. At high $p_T$, comparisons been the
jet quenching theory and the experimental measurement can (and should) 
also be made for the differential cross sections. At low $p_T$, a 
deviation is present due to the fixed order baseline pQCD
calculation and QGP effects are more accurately studied via
$R_{AA}(p_T)$.

\section{ Quenching of inclusive pions at RHIC and the LHC }
\label{result}

\begin{figure}[!t]
\includegraphics[width=3.2in,height=3.9in,angle=0]{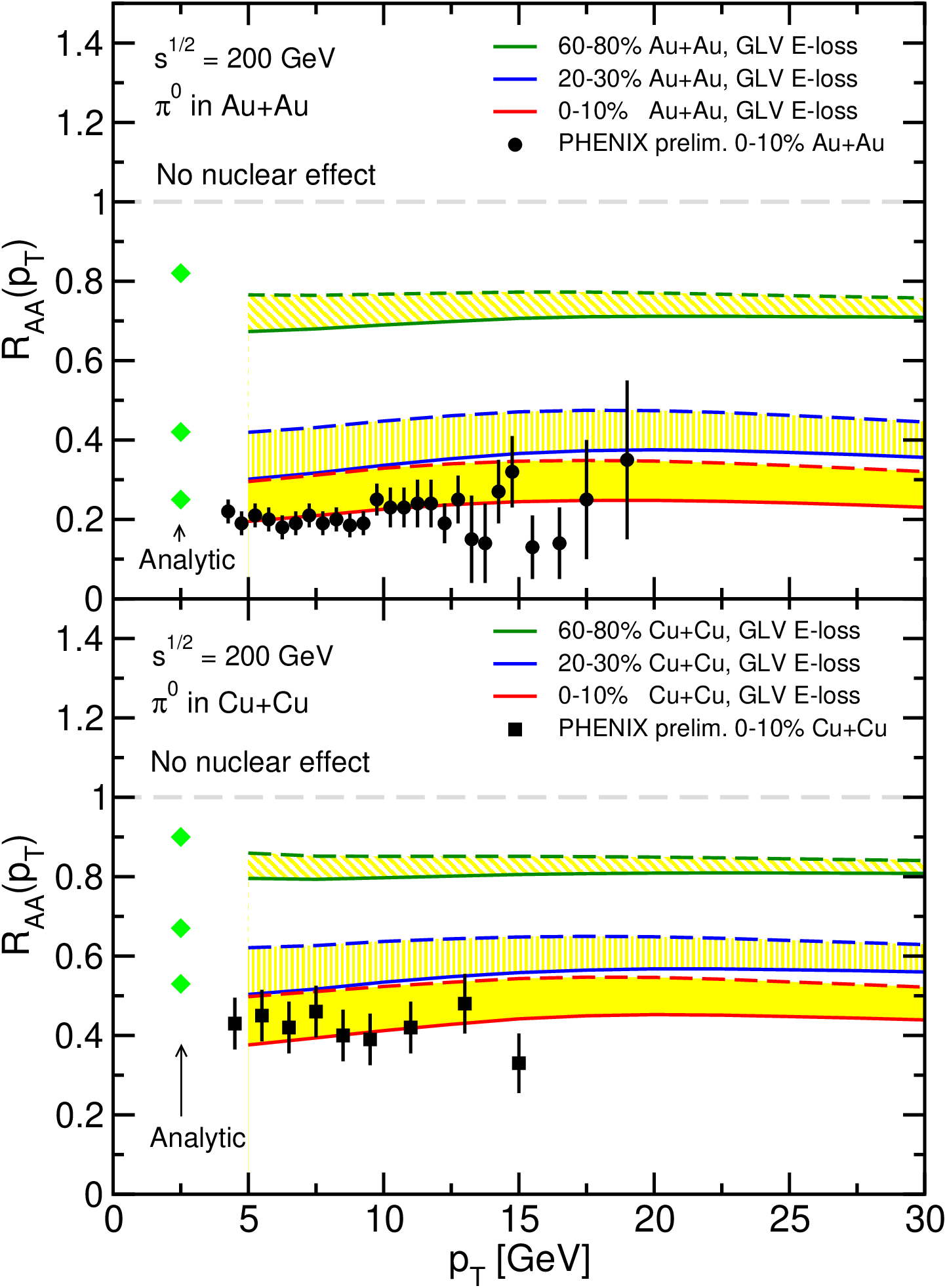}
\caption{ Top panel: nuclear modification factor $R_{AA}$ for Au+Au
collisions at $\sqrt{s} = 200$~GeV versus $p_T$ and centrality.  
Bottom panel: similar calculation for Cu+Cu at $\sqrt{s} = 200$~GeV 
at RHIC. Preliminary data is from PHENIX~\cite{Shimomura:2005en} and
analytic estimates from Fig.~\ref{analytic} are also shown. }
\label{cros-rel}
\end{figure}

\begin{figure}[!t]
\includegraphics[width=3.2in,height=2.8in,angle=0]{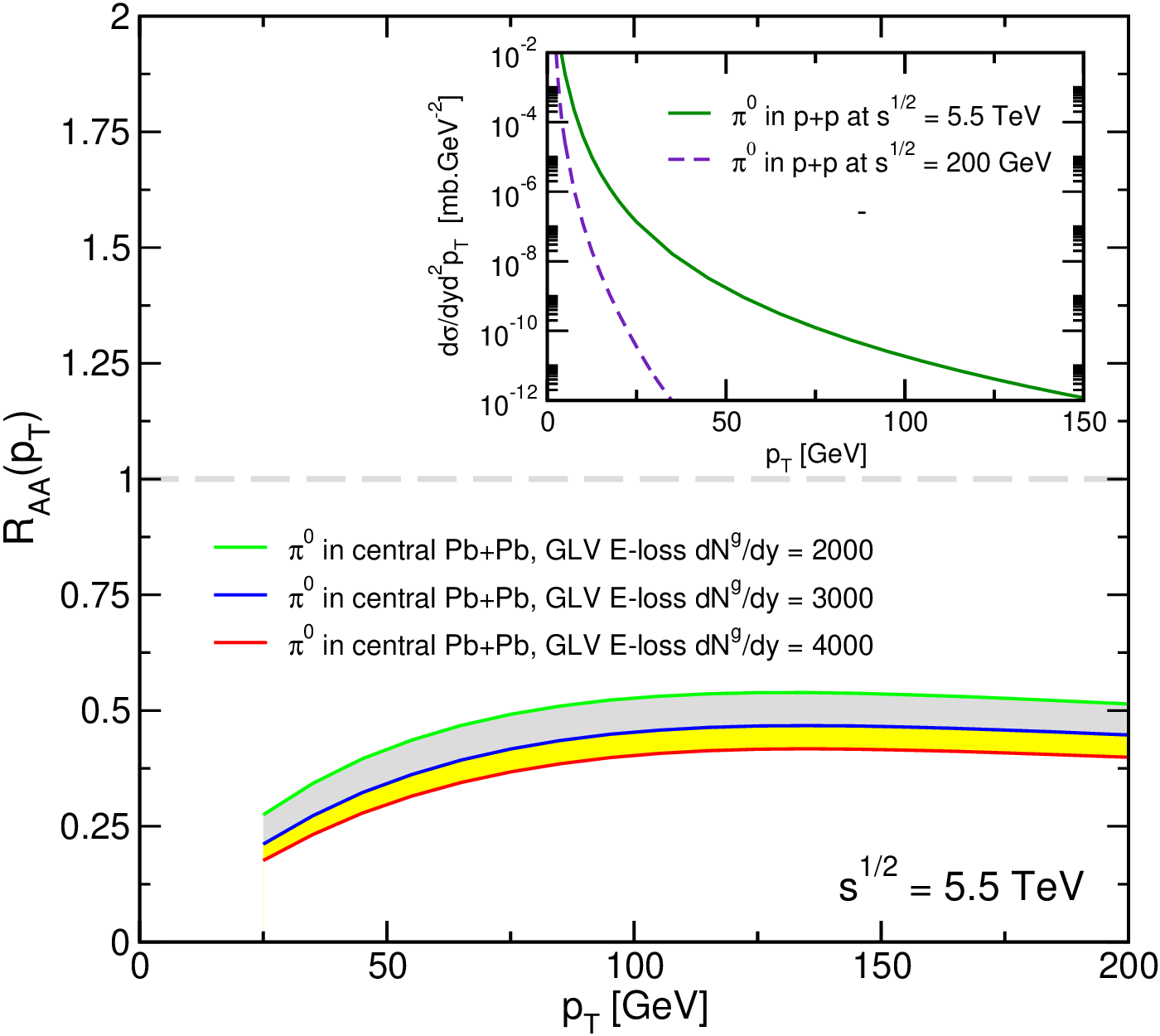}
\caption{ Suppression of $\pi^0$ production in central Au+Au 
collisions at the LHC as a function of the parton rapidity 
density. Insert shows the baseline p+p $\pi^0$ cross section 
at $\sqrt{s}=200$~GeV and $\sqrt{s} = 5.5$~TeV. }
\label{cros-LHC}
\end{figure}

Having evaluated the energy loss of quark and gluon jets
in the QGP media specified in Table~\ref{table-AuCu}, we
calculate the quenched pion spectra using Eqs.~(\ref{single}),
(\ref{isoemc}), (\ref{cronin}) and (\ref{nucmod-1}). 
Figure~\ref{cros-rel} shows $R_{AA}(p_T)$ for central,
semi-central and peripheral collisions.  
The predictions in Cu+Cu reactions are for a constant 
suppression factor, as in Au+Au, at high $p_T$. 
Preliminary  PHENIX data, first compared to this theory 
in Ref.~\cite{Shimomura:2005en}, are also included.  It should 
be noted that even in peripheral reactions there can be 
noticeable particle attenuation. This is larger 
than the analytic estimates due to  sub-leading 
effects of the medium density on the parton energy loss, 
discussed in the previous  Section, see Fig.~\ref{cros-rel}.
Whether such effects are observable or compensated by the
non-uniform QGP density in the transverse plane  is 
subject to experimental verification. Finally, 
we make the important observation that for similar densities 
and system sizes, for  example $dN^g/dy = 410$ in mid-central 
Au+Au and $dN^g/dy = 370$ in central Cu+Cu, the magnitude 
of the predicted pion suppression is similar.

Pb+Pb collisions at the LHC represent the future energy 
frontier of QGP studies in heavy ion reactions. We have 
explored the sensitivity of $R_{AA}(p_T)$ to 
the parton rapidity density in central nuclear reactions with 
$dN^g/dy \simeq 2000, 3000$ and $4000$.  In~\cite{Turbide:2005fk}
a seven-fold increase of the medium density in going form RHIC to 
the LHC was assumed. In this work we adhere to a more modest 
two- to  four-fold increase of the soft hadron rapidity 
density and emphasize that future measurements of jet 
quenching must be  
correlated to  $dN^g/dy \simeq dN^{h}/dy \approx (3/2) 
dN^{ch}/dy$~\cite{Vitev:2002pf,Adil:2004cn,Gyulassy:2000gk}
to verify the consistency of the phenomenological results.

\begin{figure}[t!]
\includegraphics[width=3.in,height=3.8in,angle=0]{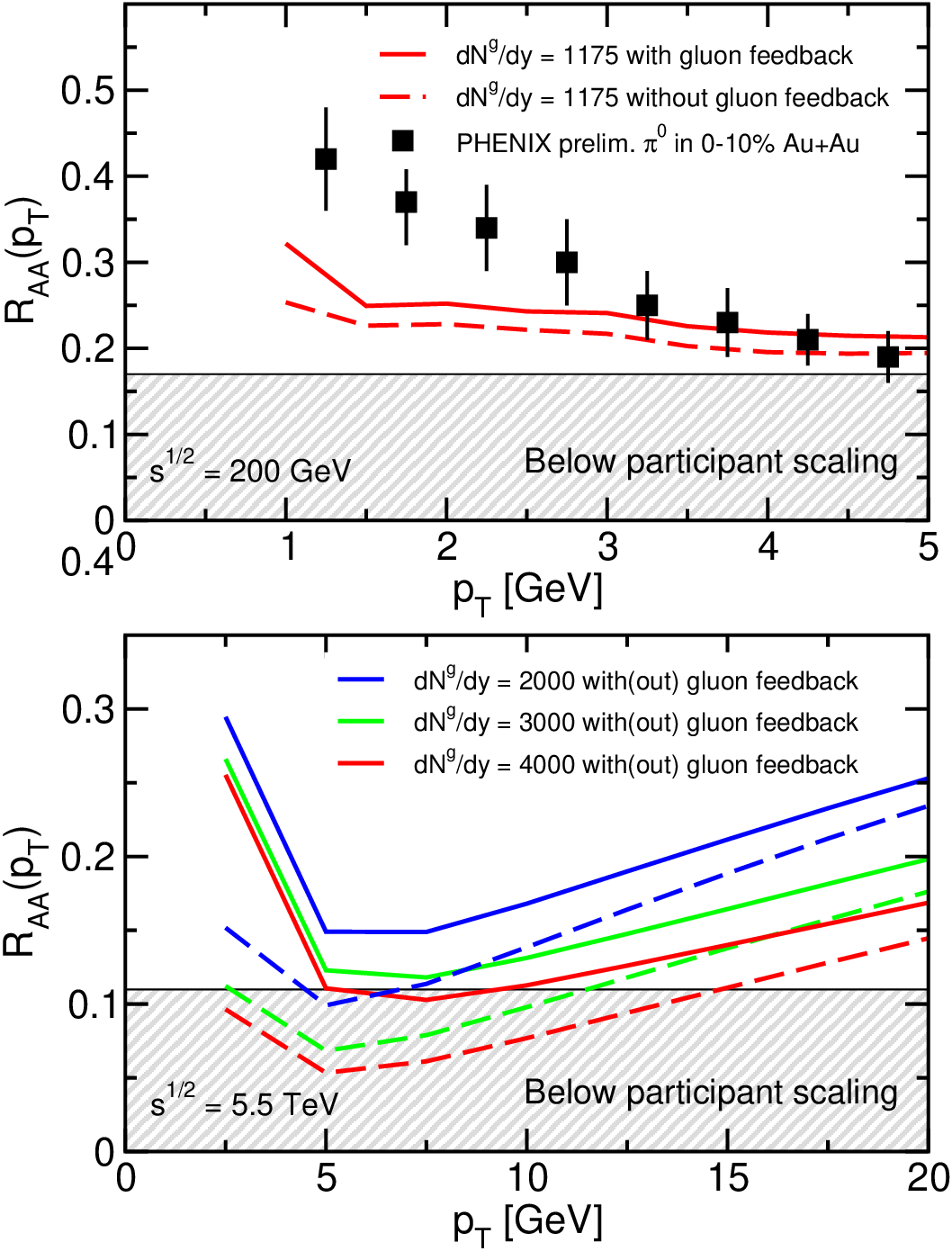}
\caption{ Top panel: nuclear modification factor $R_{AA}$ in 0-10\% 
central Au+Au collisions at moderate $p_T$ with (solid line) and without 
(dashed line) gluon feedback, $dN^g/dy = 1175$. Preliminary data 
is from PHENIX~\cite{Shimomura:2005en}. 
Bottom panel: nuclear suppression at moderate $p_T$ at the LHC 
$\sqrt{s}=5.5$~TeV. Central Pb+Pb collisions
with (solid line) and without (dashed line) gluon feedback are shown,  
$dN^g/dy \simeq 2000, 3000, 4000$.}
\label{cros-feed}
\end{figure}

From Figs.~\ref{cros-rel},~\ref{cros-LHC} and~\ref{cros-feed}  
we conclude that
at low and moderate $p_T$, jet suppression at the LHC 
is larger than at RHIC. However, at high $p_T > 30 - 50$~GeV 
this ordering is reversed.
The physics reason for our result 
is that the {\em fractional} energy loss 
$\langle \,  \Delta E / E \, \rangle $  
of  100~GeV quark and gluon jets at the LHC is not very 
different from that of 25~GeV jets at RHIC,  see Fig.~\ref{ElossNg}.  
In addition, the power law dependence of the hadronic and the 
underlying partonic spectrum $n \, (y,p_{T_c},\sqrt{s})$ changes
(decreases)  from $\sqrt{s} = 200$~GeV to  $\sqrt{s} =  5.5$~TeV. 
This is shown 
in the insert of Fig.~\ref{cros-LHC} and affects the 
magnitude of calculated nuclear suppression, 
Eq.~(\ref{raa-analyt}). 
LHC will soon provide an  extended $p_T$ range to test
the correlation of $R_{AA}(p_T)$ with the stiffness 
of the differential particle spectra.

To assess the importance of the gluon feedback term 
in Eq.~(\ref{nucmod-1}), we extend, in Fig.~\ref{cros-feed},   
the calculation of $R_{AA}(p_T)$ for central Au+Au and Pb+Pb 
collisions to low and moderate transverse momenta. At RHIC the
redistribution of the lost energy leads to small, $\sim 25\%$, 
modification of the neutral pion cross section in the 
$p_T \sim 1 - 2 $ GeV range and modest improvement in the 
theoretical description of that data.  
In contrast, at the LHC the fragmentation  of medium-induced gluons
is a much more significant $\geq 100\% $ correction to the  low 
and moderate $p_T$  $\pi^0$ production rate.

The need for the more consistent treatment of jet energy 
loss,  Eq.~(\ref{nucmod-1}), is also illustrated by 
comparing  $R_{AA}(p_T)$  in Fig.~\ref{cros-feed}  to  the  
participant  scaling ratio: $(N_{\rm part}/2)/N_{\rm col}$~\cite{part}.
At high $p_T$ there is no lower limit on the quenching 
of jets. At low $p_T$ the total available energy 
of the collision, $ \sim  (N_{\rm part} /2) \sqrt{s} $,  
suggests participant scaling of bulk particle production 
confirmed by hydrodynamic calculations~\cite{Hirano:2003pw}.
Previous estimates of leading particle suppression 
at low and moderate $p_T$  have violated this 
limit~\cite{Vitev:2002pf,Turbide:2005fk}.  
Fig.~\ref{cros-feed} shows that the gluon feedback can ensure
numerically  $R_{AA}(p_T) \geq (N_{\rm part}/2) / N_{\rm col}$ at 
mid-rapidity at the LHC for the densities considered here
and is important everywhere in the region $ p_T \leq 15$~GeV.

\section{Conclusions}

In this Letter we  presented predictions for the nuclear 
modification of inclusive neutral pion production 
in Au+Au and Cu+Cu collisions at $\sqrt{s} = 200$~GeV versus
centrality and transverse momentum. Pb+Pb reactions 
at $\sqrt{s} = 5.5$~TeV at the LHC were also discussed 
and our calculations at mid-rapidity 
accounted for Cronin multiple 
scattering~\cite{Adams:2003im}, nuclear shadowing~\cite{Eskola:1998df}  
and final state radiative energy loss in the quark-gluon 
plasma~\cite{Gyulassy:2000fs}. Elastic energy loss was not 
considered here, since its effects are still under 
debate~\cite{Peigne:2005rk} relative to the attenuation 
of jets via gluon bremsstrahlung. 
We compared our numerical results to a simplified analytic model
for centrality dependent jet quenching~\cite{Vitev:2002pf} 
and showed that there are non-negligible corrections 
in the evaluation and implementation of radiative energy 
loss related to the temperature or $\mu$ dependence of the 
strong coupling constant. While the dependence of the 
observable hadron suppression on $dN^g/dy$  was shown to be 
sub-linear, this calculation retains sensitivity to both the
properties of the medium and the underlying perturbative 
baseline cross sections.

At low and moderate transverse momenta we derived 
the contribution to single inclusive pion production 
from the fragmentation of medium-induced gluons. 
At RHIC we found this effect to be a modest, $\leq 25\% $, 
correction. Our result should be contrasted with the case of 
back-to-back  di-hadron correlations where the redistribution 
of the lost energy controls the QGP-induced transition from 
suppression to enhancement of large angle inclusive two 
particle production~\cite{Vitev:2005yg}. At the 
LHC, however, even in inclusive one pion
calculations, gluon feedback is shown to alter the 
$p_T \leq 15$~GeV cross section in central 
Pb+Pb reactions by as much as a factor of two. In summary, 
results reported in this Letter not only provide a more 
consistent theoretical framework to treat the effects
of medium-induced gluon bremsstrahlung but also rectify 
the over-quenching of jets at low $p_T$ in the limit of large 
fractional energy loss~\cite{Vitev:2002pf,Turbide:2005fk}.

\vspace*{0.5cm}

\begin{acknowledgments}
This research is supported in part by the US Department of Energy  
under Contract No. W-7405-ENG-3 and by 
the J. Robert Oppenheimer Fellowship of the Los Alamos National 
Laboratory. 
\end{acknowledgments}


\begin{thebibliography}{99}


\bibitem{Shimomura:2005en}
  M.~Shimomura  [PHENIX Collaboration],
nucl-ex/0510023; 
  B.~Cole  [PHENIX Collaboration], Quark Matter 2005 proceedings.


\bibitem{Dunlop:2005xe}
  J.~C.~Dunlop [STAR Collaboration],
nucl-ex/0510073;
  P.~M.~Jacobs and M.~van Leeuwen [STAR Collaboration],
nucl-ex/0511013.


\bibitem{Gyulassy:2004vg}
  M.~Gyulassy,
  nucl-th/0403032;
  I.~Vitev,
  J.\ Phys.\ G {\bf 30}, S791 (2004).


\bibitem{Levai:2001dc}
P.~Levai {\it et al.}, 
Nucl.\ Phys.\ A {\bf 698}, 631 (2002); 
K.~Adcox {\it et al.}, 
Phys.\ Rev.\ Lett.\  {\bf 88}, 022301 (2002).


\bibitem{Adams:2003im}
  S.~S.~Adler {\it et al.}  [PHENIX Collaboration],
  Phys.\ Rev.\ Lett.\  {\bf 91}, 072303 (2003);
  J.~Adams {\it et al.}  [STAR Collaboration],
  Phys.\ Rev.\ Lett.\  {\bf 91}, 072304 (2003);
  I.~Vitev, 
  Phys.\ Lett.\ B {\bf 562}, 36 (2003).


\bibitem{Vitev:2002pf}
  I.~Vitev and M.~Gyulassy,
  Phys.\ Rev.\ Lett.\  {\bf 89}, 252301 (2002);
  X.~N.~Wang,
  Phys.\ Rev.\ C {\bf 70}, 031901 (2004).


\bibitem{Adler:2003qi}
S.~S.~Adler {\it et al.}, 
Phys.\ Rev.\ Lett.\  {\bf 91}, 072301 (2003);
  J.~Adams {\it et al.}  [STAR Collaboration],
  Phys.\ Rev.\ Lett.\  {\bf 91}, 172302 (2003).


\bibitem{Adil:2004cn}
  A.~Adil and M.~Gyulassy,
  Phys.\ Lett.\ B {\bf 602}, 52 (2004);
  I.~Vitev,
  Phys.\ Lett.\ B {\bf 606}, 303 (2005).


\bibitem{Turbide:2005fk}
  S.~Turbide, C.~Gale, S.~Jeon and G.~D.~Moore,
  Phys.\ Rev.\ C {\bf 72}, 014906 (2005);
  A.~Dainese, C.~Loizides and G.~Paic,
  Eur.\ Phys.\ J.\ C {\bf 38}, 461 (2005).


\bibitem{Gyulassy:2000fs}
M.~Gyulassy, P.~Levai,  I.~Vitev,
Phys.\ Rev.\ Lett.\  {\bf 85}, 5535 (2000);
Nucl.\ Phys.\ B {\bf 594}, 371 (2001).




\bibitem{Vitev:2005yg}
  I.~Vitev,
  Phys.\ Lett.\ B {\bf 630}, 78 (2005); 
  J.~Adams {\it et al.}  [STAR Collaboration],
  Phys.\ Rev.\ Lett.\  {\bf 95}, 152301 (2005).


\bibitem{Gyulassy:2000gk}
  M.~Gyulassy, I.~Vitev and X.~N.~Wang,
  Phys.\ Rev.\ Lett.\  {\bf 86}, 2537 (2001).


\bibitem{Arleo:2002kh}
  F.~Arleo,
  JHEP {\bf 0211}, 044 (2002);
  M.~Gyulassy, P.~Levai and I.~Vitev,
  Phys.\ Lett.\ B {\bf 538}, 282 (2002).




\bibitem{Eskola:1998df}
  K.~J.~Eskola, V.~J.~Kolhinen and C.~A.~Salgado,
  Eur.\ Phys.\ J.\ C {\bf 9}, 61 (1999);
  J.~W.~Qiu and I.~Vitev,
  Phys.\ Lett.\ B {\bf 632}, 507 (2006).


\bibitem{part} With $\sigma_{pp}^{in} = 42$~mb  
  (65~mb)  we obtain $(N_{\rm part}/2)/N_{\rm col} = 0.17 \; (0.11)$ 
at RHIC (LHC), respectively.


\bibitem{Hirano:2003pw}
  T.~Hirano and Y.~Nara,
  Phys.\ Rev.\ C {\bf 69}, 034908 (2004).



\bibitem{Peigne:2005rk}
  S.~Peigne, P.~B.~Gossiaux and T.~Gousset,
 JHEP {\bf 0604}, 011 (2006);
  S.~Wicks, W.~Horowitz, M.~Djordjevic and M.~Gyulassy,
  nucl-th/0512076.




\end{thebibliography}
\end{document}